\documentclass[%
 aip,
 amsmath,amssymb,
 reprint,%
]{revtex4-1}

\usepackage{graphicx}
\usepackage{dcolumn}
\usepackage{bm}

\usepackage[utf8]{inputenc}
\usepackage[T1]{fontenc}
\usepackage{mathptmx}
\usepackage{etoolbox}

\usepackage{amsmath,amssymb}
\usepackage{graphicx}
\usepackage{dcolumn}
\usepackage{bm}
\usepackage{float}
\usepackage[dvipsnames]{xcolor}
\usepackage{hyperref}
\hypersetup{colorlinks=true,allcolors=MidnightBlue}
\usepackage{braket}  
\usepackage{xcolor}

\makeatletter
\def\@email#1#2{%
 \endgroup
 \patchcmd{\titleblock@produce}
  {\frontmatter@RRAPformat}
  {\frontmatter@RRAPformat{\produce@RRAP{*#1\href{mailto:#2}{#2}}}\frontmatter@RRAPformat}
  {}{}
}%
\makeatother
\begin{document}

\preprint{AIP/123-QED}

\title{The Convexity Condition of Density-Functional Theory}
\author{Andrew C. Burgess}
 \affiliation{School of Physics, Trinity College Dublin, The University of Dublin, Ireland}
\author{Edward Linscott}%
\affiliation{ 
Theory and Simulation of Materials (THEOS), École Polytechnique Fédérale de Lausanne, 1015 Lausanne, Switzerland 
}%

\author{David D. O'Regan}
 \email{david.o.regan@tcd.ie}
\affiliation{School of Physics, Trinity College Dublin, The University of Dublin, Ireland}%

\date{\today}

\begin{abstract}
    It has long been postulated that within density-functional theory (DFT) the total energy of a finite electronic system is convex with respect to electron count, so that $2E_v[N_0] \leq E_v[N_0-1]+E_v[N_0+1]$. Using the infinite-separation-limit technique, this article proves the convexity condition for any formulation of DFT that is (1) exact for all $v-$representable densities, (2) size-consistent, and (3) translationally invariant. An analogous result is also proven for one-body reduced density matrix functional theory. While there are known DFT formulations in which the ground state is not always accessible, indicating that convexity does not hold in such cases, 
    this proof nonetheless confirms a stringent constraint on the exact exchange-correlation functional. We also provide sufficient conditions for convexity in approximate DFT, which could aid in the development of density-functional approximations. This result lifts a standing assumption in the proof of the piecewise linearity condition with respect to electron count, which has proven central to understanding the Kohn-Sham band-gap and the exchange-correlation derivative discontinuity of DFT.
\end{abstract}

\maketitle

To acknowledge his deep and numerous contributions to Density-Functional Theory (DFT), it is a pleasure to dedicate this 
article to John Perdew on the occasion of his eightieth birthday.
With over $240,000$ publications on the Web of Science to date~\cite{vannoordenTop100Papers2014,haunschildComprehensiveAnalysisHistory2019}, the DFT approach to quantum 
mechanics has provided an extremely popular tool for computational chemistry and condensed matter physics~\cite{hohenbergInhomogeneousElectronGas1964a,kohnSelfConsistentEquationsIncluding1965,m.tealeDFTExchangeSharing2022,perdewFourteenEasyLessons2010}. DFT owes much of its success to the development of relatively efficient, reliable and accurate Density-Functional Approximations (DFAs)~\cite{voskoAccurateSpindependentElectron1980a,perdewGeneralizedGradientApproximation1996a,beckeDensityfunctionalExchangeenergyApproximation1988a,beckeDensityFunctionalThermochemistry1993a,heydHybridFunctionalsBased2003a,leeDevelopmentColleSalvettiCorrelationenergy1988a, perdewGeneralizedGradientApproximation1996b,perdewAccurateSimpleAnalytic1992}. Although the explicit form of the exact density functional remains unknown, there exists a number of known exact physical conditions that it must obey~\cite{kaplan2023predictive}, such as the piecewise linearity conditions in electron count~\cite{perdewDensityFunctionalTheoryFractional1982,yangDegenerateGroundStates2000a,ayersDependenceContinuityEnergy2008} and magnetization~\cite{burgessTiltedplaneStructureEnergy2023,galEnergySurfaceChemical2010,cohenFractionalSpinsStatic2008}, the Lieb-Oxford bound~\cite{liebImprovedLowerBound1981,kin-licchanOptimizedLiebOxfordBound1999,lewinImprovedLiebOxford2022}, and the virial theorem~\cite{lowdinScalingProblemVirial1959,levyHellmannFeynmanVirialScaling1985,averillVirialTheoremDensityfunctional1981,cruzExchangeCorrelationEnergy1998}. DFAs are typically designed to enforce a selected subset of these known conditions~\cite{perdewGeneralizedGradientApproximation1996a,sunStronglyConstrainedAppropriately2015,perdewDensityFunctionalFull2008,hendersonRangeSeparationLocal2008,maierAssessingLocallyRangeseparated2021,vuckovicDensityFunctionalsBased2023,perdewPrescriptionDesignSelection2005}, even in the case of some machine-learned functionals~\cite{kirkpatrickPushingFrontiersDensity2021,nagaiMachinelearningbasedExchangeCorrelation2022,hollingsworthCanExactConditions2018,pokharelExactConstraintsAppropriate2022}. The elucidation of such conditions is thus of paramount importance to improve the performance of practical DFT calculations. 

It is often {\it assumed} that another such exact condition is the convexity condition with respect to electron count $N$, which states that, for any finite $N_0$-electron system, 
\begin{equation}
\label{eqn:convexity_condition}
2E_v[N_0] \leq E_v[N_0-1]+E_v[N_0+1] \quad N_0, \in \mathbb{N}^0.
\end{equation}
This assumption is based on the lack of any experimental counterexample~\cite{yangCommunicationTwoTypes2016}, yet a rigorous proof has been lacking to date~\cite{dreizlerDensityFunctionalTheory2012,gouldAsymptoticBehaviorHartreeexchange2019}. Indeed, to quote Parr and Yang in Ref.~\onlinecite{parrDensityFunctionalTheoryAtoms}, "{\it For atoms and molecules, no counterexample is known  \ldots although a first-principles proof has never been given}". 
The intended purpose of the present work is to provide such a proof for all formulations of DFT that satisfy certain minimal requirements. For the purposes of this study, we therefore define the ground-state energy of the $N$ electron system with external potential $v({\bf r})$,
\begin{equation}
\label{eqn:ground_state_minimization1}
E_{v}[N]=  \min_{\substack{\rho \rightarrow \int  \rho(\mathbf{r}
) d\mathbf{r}=N}}E_{v}[\rho({\bf r})].
\end{equation}
Yang et al.~\cite{yangDegenerateGroundStates2000a} have shown that in the zero-temperature grand canonical ensemble formulation of DFT, defined for an arbitrary, possibly fractional number of electrons, the total energy must obey the convexity condition (see also G{\'a}l et al.~\cite{galEnergySurfaceChemical2010}). Ayers et al.~\cite{ayersLevyConstrainedSearch} have shown that this condition is also satisfied in the Fock space constrained search formulation of DFT. In this article, we extend the work of Yang, Ayers et al. and show that any formulation of DFT that is (1) exact for all $v-$representable densities, (2) size-consistent, and (3) translationally invariant must obey the convexity condition.  We anticipate, but do not assume,  that the exact functional should satisfy conditions (1) to (3) in any DFT formulation
where the ground state is generally accessible. Before explicating this proof of the convexity condition we will first emphasize that numerous different formulations of DFT exist. 

In 1964, through their famous proof by contradiction, Hohenberg and Kohn (HK)~\cite{hohenbergInhomogeneousElectronGas1964a} offered the first formulation of DFT,
\begin{equation}
E_{\rm HK}[\rho_v]=\int d{\bf r} \enskip \rho_v({\bf r})v({\bf r}) +F_{\rm HK}[\rho_v],
\end{equation}
where $F_{\rm HK}[\rho_v]$ is the HK universal functional (the sum of kinetic and electron-electron interaction energies), so-called due to its independence from the system under consideration. The notation $\rho_v({\bf r})$ is used to emphasise that the HK formulation of DFT is only defined for $v$-representable densities, i.e., the density associated with the ground-state wavefunction of some Hamiltonian --- a serious limitation when searching for ground-state densities~\cite{englischHohenbergKohnTheoremNonVrepresentable1983,ayersAxiomaticFormulationsHohenbergKohn2006,ayersNecessarySufficientConditions2007,levyElectronDensitiesSearch1982,gonisReformulationDensityFunctional2016,gunnarssonExchangeCorrelationAtoms1976}. The HK formulation of DFT was also extended to non-zero temperature by Mermin~\cite{merminThermalPropertiesInhomogeneous1965}.

Levy~\cite{levyUniversalVariationalFunctionals1979} offered an alternative formulation of DFT (see also Lieb~\cite{liebDensityFunctionalsCoulomb1983}), which is defined for all $N$-representable densities, i.e., the density associated with some antisymmetric wavefunction $\Psi$, such that 
\begin{equation}
E_{\rm Levy}[\rho]=\int d{\bf r} \enskip 
\rho({\bf r})v({\bf r})+\min_{\Psi \rightarrow \rho}\braket{\hat{T}+\hat{V}_{\rm ee}}_{\Psi},
\end{equation}
where $\hat{T}$ and $\hat{V}_{\rm ee}$ are the kinetic energy and electron-electron interaction operators. Valone~\cite{valoneOneOneMapping2008} extended the Levy constrained search formulation to all $N$-particle density operators,
\begin{equation}
\label{eqn:Valone}
E_{\rm Valone}[\rho]=\int d{\bf r} \enskip \rho({\bf r})v({\bf r})+\min_{\hat{\Gamma}_N \rightarrow \rho}\braket{\hat{T}+\hat{V}_{\rm ee}}_{\hat{\Gamma}_N},
\end{equation}
where the $N$-particle density operator is
\begin{equation}
\hat{\Gamma}_N=\sum_i p_{i}\ket{\Psi_{Ni}}\bra{\Psi_{Ni}}.
\end{equation}
This was later extended by Perdew et al.~\cite{perdewDensityFunctionalTheoryFractional1982} to a constrained search over all mixed-states, including non-integer particle counts, in which case ${\hat{\Gamma}_N}$ in Eq.~\ref{eqn:Valone} is replaced by the ensemble density operator
\begin{equation}
\hat{\Gamma}=\sum_{N}\sum_i p_{Ni}\ket{\Psi_{Ni}}\bra{\Psi_{Ni}}.
\end{equation}
This formulation of DFT is referred to as the zero-temperature grand canonical ensemble formulation.
As mentioned already, Ayers et al.~\cite{ayersLevyConstrainedSearch} extended
the Levy constrained-search formulation to all wavefunctions in Fock space
\begin{equation}
\ket{\Psi}=\sum_N c_N \ket{\Psi_N}.
\end{equation}
By construction, this yields the same ground-state total energies as the zero-temperature grand canonical ensemble formulation.

{\bf Illustrative Example} We claim that all formulations of DFT that satisfy conditions (1) to (3) must obey the convexity condition. In order to show this, suppose that there exists some $N_0$-electron finite system with external potential $v({\bf r})$, which breaks the convexity condition. Then we have
\begin{equation}
\label{eqn:convexity_violation}
2E_v[N_0] > E_v[N_0-1]+ E_v[N_0+1].
\end{equation}
In order to proceed, we employ the infinite-separation-limit technique~\cite{yangDegenerateGroundStates2000a,ayersDependenceContinuityEnergy2008} and consider the external potential $v^{\prime}({\bf r})$, which is composed of two copies of the same external potential $v({\bf r})$, infinitely separated in space, such that 
\begin{equation}
v^{\prime}({\bf r})=\sum_{l=1}^2 v_{{\bf R}_l}({\bf r}).
\end{equation}
In this illustrative example (only), we  assume that the $2N_0$-electron ground-state wavefunction $\Psi_1$ of the system, with external potential $v^{\prime}({\bf r})$, is the anti-symmetric product of $\Phi_{N_0-1}({\bf{R}}_1)$ and $\Phi_{N_0+1}({\bf{R}}_2)$, where $\Phi_{N_0-1}({\bf{R}}_1)$ and $\Phi_{N_0+1}({\bf{R}}_2)$ are the ground-state wavefunctions of the $N_0-1$- and $N_0+1$-electron systems in the external potential $v({\bf{r}})$. Specifically, we write 
\begin{equation}
\label{eqn:psi1}
\Psi_1=\hat{A}\bigg(\Phi_{N_0-1}({\bf{R}}_1)\Phi_{N_0+1}({\bf{R}}_2)\bigg).
\end{equation}
Eq.~\ref{eqn:psi1} is an assumption as one cannot rule out the possibility that the true ground-state wavefunction is for example, the anti-symmetric product of $\Phi_{N_0-2}({\bf{R}}_1)$ and $\Phi_{N_0+2}({\bf{R}}_2)$. In the complete proof, this assumption is not invoked because the site electron counts of the wavefunctions $\Phi({\bf{R}}_l)$ are not specified. We will first show that the convexity condition cannot be violated in this simple case, after which a full proof of the convexity condition will be given. 

One may construct the wavefunction $\Psi_2$ by reversing the position vectors ${\bf{R}_1}$ and ${\bf{R}_2}$,
\begin{equation}
\Psi_2=\hat{A}\bigg(\Phi_{N_0-1}({\bf{R}}_2)\Phi_{N_0+1}({\bf{R}}_1)\bigg).
\end{equation}
The averaged wavefunction
defined by
\begin{equation}
\label{eqn:averaged_wavefunction2}
\Psi_{\rm avg}=\frac{1}{\sqrt{2}}[\Psi_1+\Psi_2]
\end{equation}
will be degenerate with $\Psi_1$ and $\Psi_2$ and its corresponding density will be given by
\begin{equation}
\label{eqn:averaged_wavefunction}
\rho({\bf r})=\sum_{l=1}^2\frac{1}{2}\rho_l({\bf r};N_0-1)+\frac{1}{2}\rho_l({\bf r};N_0+1),
\end{equation}
where $\rho_l({\bf r};N_0\pm1)$ is the ground-state electron density of site $l$ with an electron count $N_0\pm1$. To deduce an expression for the total energy of the system with electron density $\rho({\bf r})$, we make three assumptions about the nature of the total energy functional that were previously discussed, namely  that it is (1) exact for all $v-$representable densities, (2) size-consistent, and (3) translationally
invariant. For the purposes of the proof, these three assumptions are treated as axioms. 

Given that $\Psi_1$ of Eq.~\ref{eqn:psi1} is a ground-state wavefunction of the system, we know that $\rho({\bf r})$ of Eq.~\ref{eqn:averaged_wavefunction}  is a ground-state density for the system with external potential $v'({\bf r})$ and is thus $v-$representable. From assumption 1 it follows that the total energy functional is exact for the electron density $\rho({\bf r})$. Its total energy should be equal to that of $\Psi_1$, which is simply the sum of the ground-state energies of two infinitely separated sites with electron counts $N_0 \pm1$,
\begin{equation}
\label{eqn:exactE}
E_{v'}[\rho({\bf r})]=E_{v_{{\bf R}_l}}[N_0-1]+E_{v_{{\bf R}_l}}[N_0+1].
\end{equation}
Here, $E_{v}[N]$ is the ground-state energy of the $N$ electron system with external potential $v({\bf r})$,
\begin{equation}
\label{eqn:ground_state_minimization}
E_{v}[N]=  \min_{\substack{\rho \rightarrow \int  \rho(\mathbf{r}
) d\mathbf{r}=N}}E_{v}[\rho({\bf r})].
\end{equation}
From assumption 2, the total energy functional should be size-consistent, whereby
\begin{equation}
\label{eqn:sizeconsistentE}
E_{v'}[\rho({\bf r})]=\sum_{l=1}^{2}E_{v_{{\bf R}_l}}\bigg[\frac{1}{2}\rho_l({\bf r};N_0-1)+\frac{1}{2}\rho_l({\bf r};N_0+1)\bigg].
\end{equation}
Eq.~\ref{eqn:sizeconsistentE} can be simplified by application of the translational invariance assumption to give
\begin{equation}
\label{eqn:translationalinvarianceE}
E_{v'}[\rho({\bf r})]=2E_{v_{{\bf R}_l}}\bigg[\frac{1}{2}\rho_l({\bf r};N_0-1)+\frac{1}{2}\rho_l({\bf r};N_0+1)\bigg],
\end{equation}
where the site electron count is
\begin{equation}
N=\frac{1}{2}(N_0-1)+\frac{1}{2}(N_0+1)=N_0.
\end{equation}
Eq.~\ref{eqn:translationalinvarianceE} enforces an upper bound on the total energy of the $N_0$ electron system in external potential $v_{{\bf R}_l}({\bf r})$, namely
\begin{equation}
\label{eqn:energybound}
E_{v'}[\rho({\bf r})] \geq 2 E_{v_{{\bf R}_l}}[N_0].
\end{equation}
Equality in Eq.~\ref{eqn:energybound} is achieved if there exists no wavefunction with a lower total energy per site, for any external potential composed of more than two copies of $v({\bf r})$ that are infinitely separated. Combining Eqs. \ref{eqn:exactE} and \ref{eqn:energybound} and suppressing the site index label ${\bf R}_l$, we find that
\begin{equation}
\label{eqn:inconsistency}
2E_{v}[N_0]\leq E_{v}[N_0-1]+E_{v}[N_0+1].
\end{equation}
This is in direct contradiction to our original assumption in Eq.~\ref{eqn:convexity_violation}. We now proceed to give a complete proof by contradiction of the convexity condition in the general case.

{\bf Complete Proof} Suppose there exists a sequence of consecutive integer values of electron count in the range ($N_0$, $N_0+z_1$) that break the convexity condition
\begin{equation}
\label{eqn:convexity_violation2}
2E_v[N_0+z_2]>E_v[N_0+z_2-1]+E_v[N_0+z_2+1],
\end{equation}
where $N_0$, $z_1$ and $z_2 \in \mathbb{N}^0$ and $z_2$ satisfies $0<z_2<z_1$. One may construct an external potential $v^{\prime}({\bf r})$ that is composed of $q$ copies of the same external potential $v({\bf r})$, infinitely separated in space, per
\begin{equation}
\label{eqn:multisite_external_potential}
v^{\prime}({\bf r})=\sum_{l=1}^{q} v_{{\bf R}_l}({\bf r}).
\end{equation}
The ground-state of the $q(N_0+z_2)$ electron system with external potential $v^{\prime}({\bf r})$  will be composed of $q_1$ sites with electron count $N_0-z_3$ and $q_2$ sites with electron count $N_0+z_4$, where $0 \leq z_3 \leq N_0$, $z_1\leq z_4$ and $q_1, q_2, z_3$ and $z_4 \in \mathbb{N}^0$. The value of $q_2$ is constrained such that
\begin{equation}
\label{eqn:q2-expression}
q=q_1+q_2
\end{equation}
and the value of $q_1$ is constrained so that the total electron count
satisfies 
\begin{equation}
\label{eqn:q1-expression}
q(N_0+z_2)=q_1(N_0-z_3)+(q-q_1)(N_0+z_4).
\end{equation}
This leaves only two free variables,  $z_3$ and $z_4$. The exact values of $z_3$ and $z_4$ are those that minimize the total energy of the $q(N_0+z_2)$ electron system. In constructing  the external potential $v^{\prime}({\bf r})$, the total number of sites $q$ is chosen so that the ground-state energy per site is minimized. 

One possible ground-state is for the first $q_1$ sites to have an electron count of $N_0-z_3$ and wave function $\Phi_{N_0-z_3}$, while the remaining $q-q_1$ sites  have an electron count $N_0+z_4$ and wave function $\Phi_{N_0+z_4}$. The ground-state wave function of the total system 
is then given by 
\begin{align}
\label{eqn:psi1-general}
\Psi_1=\hat{A}\bigg[\bigg(\prod_{i=1}^{q_1}&\Phi_{N_0-z_3}({\bf{R}}_i)\bigg) \bigg(\prod_{j=q_1+1}^{q} \Phi_{N_0+z_4}({\bf{R}}_{j})\bigg)\bigg].
\end{align}
Swapping position vectors ${\bf{R}}_i$ and ${\bf{R}}_j$ will result in a degenerate ground-state wavefunction. Analogously to Eq.~\ref{eqn:averaged_wavefunction2},  one may construct the averaged wavefunction $\Psi_{\rm avg}$ from a linear combination of the ground-state wavefunctions of the form given by Eq.~\ref{eqn:psi1-general}. The electron density of $\Psi_{\rm avg}$ is 
\begin{equation}
\label{eqn:density_of_psi_vg}
\rho({\bf r})=\sum_{l=1}^q\frac{q_1}{q}\rho_l({\bf r};N_0-z_3)+\frac{q_2}{q}\rho_l({\bf r};N_0+z_4),
\end{equation}
Application of the same three assumptions about the nature of the total energy functional, namely  that it is (1) exact for all $v-$representable densities, (2) size-consistent, and (3) translationally
invariant results in the total energy, of site
$l$ with an electron count $N_0+z_2$,
given by
\begin{equation}
\label{eqn:energy-expression-with-q1}
E_{v_{{\bf R}_l}}[N_0+z_2]=\frac{q_1}{q}E_{v_{{\bf R}_l}}[N_0-z_3]+\frac{q_2}{q}E_{v_{{\bf R}_l}}[N_0+z_4].
\end{equation}
Eq.~\ref{eqn:energy-expression-with-q1} can be simplified, using Eq.~\ref{eqn:q2-expression} and Eq.~\ref{eqn:q1-expression}, to eliminate $q_1$ and $q_2$, to wit
\begin{equation}
\label{eqn:linear-energy-expression}
E_{v}[N_0+z_2]=\frac{z_4-z_2}{z_3+z_4}E_{v}[N_0-z_3]+\frac{z_3+z_2}{z_3+z_4}E_{v}[N_0+z_4],
\end{equation}
where the site index label ${\bf R}_l$ has been suppressed for clarity. This expression for the total energy will hold for all integer values of electron count in the range $[N_0-z_3,N_0+z_4]$. Noting that the more restricted interval 
$[N_0,N_0+z_1]
\subseteq [N_0-z_3,N_0+z_4]$,  from Eq.~\ref{eqn:linear-energy-expression} it follows that, for all integer values of $z_2$ in the range $0< z_2< z_1$, we have 
\begin{equation}
\label{eqn:convexity_not_violated}
2E_v[N_0+z_2]=E_v[N_0+z_2-1]+E_v[N_0+z_2+1].
\end{equation}
Comparing Eq.~\ref{eqn:convexity_violation2} with Eq.~\ref{eqn:convexity_not_violated} we arrive at a contradiction. Therefore, the convexity condition as given by Eq.~\ref{eqn:convexity_condition} must hold for all finite electronic systems.

{\bf Extension to 1-RDM Functional Theory}
The complete proof of the convexity condition of DFT is also valid for any formulation of one-body reduced density matrix (1-RDM) functional theory that is (1) exact for all $v-$representable 1-RDMs, (2) size-consistent, and (3) translationally invariant. In this case, the electron density in Eq.~\ref{eqn:density_of_psi_vg} must be substituted for the 1-RDM of $\Psi_{\rm avg}$,
\begin{equation}
\gamma({\bf x},{\bf x}^{\prime})=\sum_{l=1}^q\frac{q_1}{q}\gamma_l({\bf x},{\bf x}^{\prime};N_0-z_3)+\frac{q_2}{q}\gamma_l({\bf x},{\bf x}^{\prime};N_0+z_4),
\end{equation}
where ${\bf x}$ and ${\bf x}^{\prime}$ are combined space-spin coordinates and $\gamma_l({\bf x},{\bf x}^{\prime};N)$ is the ground-state 1-RDM of site $l$ with electron count $N$. All other equations in the complete proof remain unchanged.

{\bf Extension to Approximate Density Functionals} 
Approximate density functionals will typically not be exact for all $v$-representable densities (condition 1), however it is possible to substitute this condition with two weaker conditions. An approximate density functional that satisfies (1b) the constancy condition~\cite{yangDegenerateGroundStates2000a,cohenFractionalSpinsStatic2008,burgessTiltedplaneStructureEnergy2023,jacobSpinDensityfunctionalTheory2012,burtonVariationsHartreeFock2021}, 
(1c) the density-size-consistency condition, 
(2) the functional-size-consistency condition, and (3) the translational invariance condition, will also obey the convexity condition. 

The constancy condition (1b) states that given a set of $g$-fold degenerate, possibly symmetry-broken ground-state densities $\{ \rho_i({\bf r})\}$ of a system with external potential $v({\bf r})$, the total energy functional satisfies 
\begin{align}
\label{eqn:constancy_condition}
E_v\bigg[\sum_{i=1}^gC_i\rho_i({\bf r})\bigg]=&E_v[\rho_i({\bf r})], \nonumber \\ & \forall \quad 0 \leq C_i \leq 1, \quad \sum_{i=1}^gC_i=1.
\end{align}
We note that the vast majority of existing DFAs do not satisfy this condition~\cite{cohenFractionalSpinsStatic2008}, which yields strong constraints on the exact density functional~\cite{levyKineticElectronelectronEnergies2014,ayersTightConstraintsExchangecorrelation2014}.

The functional-size-consistency condition (2) and translational invariance condition (3) are the same two conditions used in the general proof above. To avoid confusion with the density-size-consistency condition, we emphasize that the functional-size-consistency condition states that for any system with external potential $v^{\prime}({\bf r})$ composed of $q$ infinitely separated site external potentials $v_{{\bf R}_l}({\bf r})$, as given in Eq.~\ref{eqn:multisite_external_potential}, with total electron density $\rho({\bf r})$ and site electron density $\rho_l({\bf r})$, the total energy functional may be written as  
\begin{equation}
E_v[\rho({\bf r})]=\sum_{l=1}^q E_{v_{{\bf R}_l}}[\rho_l({\bf r})].
\end{equation}
The density-size-consistency condition 
(1c), however, states that for any system with external potential $v^{\prime}({\bf r})$ of the form given in Eq.~\ref{eqn:multisite_external_potential}, there exists a ground-state electron density $\rho^0({\bf r})$ that is composed of a linear combination of isolated site electron densities $\rho_l^0({\bf r};N_l)$, possibly of varying site electron count $N_l$,
so that 
\begin{equation}
\rho^0({\bf r})=\sum_{l=1}^q \rho_l^0({\bf r};N_l),
\end{equation}
were $\rho_l^0({\bf r};N_l)$ is a ground-state density of the system with external potential $v_{{\bf R}_l}({\bf r})$ and integer electron count $N_l$. 

To prove that an approximate functional that satisfies conditions (1b,1c,2,3) will also obey the convexity condition, we follow the same arguments as outlined from Eqs.~\ref{eqn:convexity_violation2} to \ref{eqn:q1-expression} in the complete proof. From the two size consistency conditions (1c,2), there will exist a symmetry-broken ground-state density of the system $\rho_1({\bf r})$ with external potential $v^{\prime}({\bf r})$, where the first $q_1$ sites have an electron count of $N_0-z_3$ and the remaining $q-q_1$ sites have an electron count $N_0+z_4$, namely 
\begin{equation}
\label{eqn:rho1_sym_broken}
\rho_1({\bf r})=\sum_{l=1}^{q_1} \rho_l({\bf r};N_0-z_3)+\sum_{l=q_1+1}^{q} \rho_l({\bf r};N_0+z_4).
\end{equation}
Swapping the electron densities $\rho_l({\bf r};N_0-z_3)$ and $\rho_l({\bf r};N_0+z_4)$ at any two sites results in a degenerate ground-state electron density. From the constancy condition, one may construct the averaged electron density $\rho({\bf r})$ from a linear combination of the ground-state electron densities of the form given by Eq.~\ref{eqn:rho1_sym_broken},
\begin{equation}
\label{eqn:density_of_psi_avg_dfa}
\rho({\bf r})=\sum_{l=1}^q\frac{q_1}{q}\rho_l({\bf r};N_0-z_3)+\frac{q_2}{q}\rho_l({\bf r};N_0+z_4),
\end{equation}
which is equivalent to the electron density of Eq.~\ref{eqn:density_of_psi_vg} in the complete proof above. 

From the functional-size-consistency condition and translational invariance condition it follows that 
\begin{equation}
E_{v^{\prime}}[\rho_1({\bf r})]=q_1E_{v_{{\bf R}_l}}[N_0-z_3]+q_2E_{v_{{\bf R}_l}}[N_0+z_4].
\end{equation}
Using these same two conditions (2,3) and recalling that the total number of sites $q$ was chosen so that the total ground-state  energy per site was minimized, it follows that  
\begin{equation}
E_{v^{\prime}}[\rho({\bf r})]=qE_{v_{{\bf R}_l}}[N_0+z_2].
\end{equation}
Invoking the constancy condition
once again, $E_{v^{\prime}}[\rho_1({\bf r})]$ must be equal to $E_{v^{\prime}}[\rho({\bf r})]$, and hence
\begin{equation}
\label{eqn:linear_expression_for_etot_dfa}
E_{v_{{\bf R}_l}}[N_0+z_2]=\frac{q_1}{q}E_{v_{{\bf R}_l}}[N_0-z_3]+\frac{q_2}{q}E_{v_{{\bf R}_l}}[N_0+z_4].
\end{equation}
An analogous proof by contradiction thus follows as per the complete proof. Thus,  any approximate functional that satisfies the weaker conditions (1b,1c,2,3) must also obey the convexity condition. We emphasize that the ground-state total energy and densities given by Eqs~\ref{eqn:constancy_condition} to \ref{eqn:linear_expression_for_etot_dfa} need not be exact.

{\bf Inaccessible $N_0$-electron pure-states} 
This proof highlights an unusual characteristic of ground-state formulations of DFT that satisfy conditions (1) to (3): the convexity condition is {\it always} satisfied for such formulations of DFT irrespective of whether the pure-state convexity condition is satisfied or not. If there exists any system whose lowest energy $N_0$-electron pure-state $\ket{\Psi_{N_0}}$ does not satisfy the pure-state convexity condition
\begin{equation}
\label{eqn:wavefunction_inequality}
2E_v[\Psi_{N_0}] \leq E_v[\Psi_{N_{0}-1}] +E_v[\Psi_{N_{0}+1}], 
\end{equation}
then the total energy and electron density associated with $\ket{\Psi_{N_0}}$ is inaccessible to such formulations of ground-state DFT. These formulations of DFT will instead yield a non-pure-state ground-state density for the $N_0$-electron system. Assuming that the pure-state convexity condition is satisfied for the $N_0\pm 1$-electron systems, this density will be equal to the site electron density of a pure-state $2N_0$ electron system with two identical sites infinitely separated in space. Based on the arguments presented in the complete proof, these formulations of DFT have a well-defined total energy for such densities. The DFT ground-state total energy for the $N_0$-electron system will be equal to the average of the ground-state total energies of the $N_0-1$ and $N_0+1$-electron systems
\begin{align}
\label{eqn:energy_average}
E_v[{N_0}]=& \frac{1}{2}E_v[{N_{0}-1}] +\frac{1}{2} E_v[{N_{0}+1}] \nonumber \\
=& \frac{1}{2}E_v[\Psi_{N_{0}-1}] +\frac{1}{2} E_v[\Psi_{N_{0}+1}] \neq E_v[\Psi_{N_{0}}] , 
\end{align}
assuming that the pure $N_0\pm 1$ electron states $\ket{\Psi_{N_0\pm 1}}$ satisfy the pure-state convexity condition. Specifically, such formulations of DFT will predict a ground-state density of the $N_0$-electron system given by 
\begin{align}
\label{eqn:density_average}
\rho_{N_0}({\bf r}) &= \frac{1}{2}\rho_{N_0-1}({\bf r})+\frac{1}{2}\rho_{N_0+1}({\bf r})  \\
\nonumber
&=\frac{1}{2}\braket{\Psi_{N_{0}-1}|\hat{\rho}_{N_0-1}|\Psi_{N_{0}-1}}+\frac{1}{2}\braket{\Psi_{N_{0}+1}|\hat{\rho}_{N_0+1}|\Psi_{N_{0}+1}},
\end{align}
where $\hat{\rho}_{N_0\pm 1}$ is the $N_0 \pm 1$ electron density operator.

It is worth emphasizing~\cite{yangCommunicationTwoTypes2016} that no experiment has ever found an electronic system that violates the inequality of Eq.~\ref{eqn:wavefunction_inequality}. The pure-state convexity condition of Eq.~\ref{eqn:wavefunction_inequality} remains an open area of research in its own right. In particular, Gonis et al~\cite{gonisEnergyConvexityConsequence2014,gonisAntisymmetricWaveFunctions2011,gonisSelfentanglementDissociationHomonuclear2014} propose a proof of the $E_v[\Psi_{N_0}]$ convexity condition through a description of mixed-states as pure-states in augmented Hilbert spaces. 
Examples have been found of model systems that violate the pure-state convexity condition of Eq.~\ref{eqn:wavefunction_inequality} when the $1/|{\bf r}|$ inter-electron Coulombic potential is replaced by an alternative potential. In particular, Lieb~\cite{liebDensityFunctionalsCoulomb1983} has identified a specific two-electron system with a hard core inter-electron potential that violates the pure-state convexity condition. If it is possible to construct a density-functional for the hard core inter-electron potential that satisfies conditions (1) to (3), such a density-functional will, for this specific two-electron example, yield a ground-state energy and density equal to the average of the one and three electron pure-states as detailed in Eqs.~\ref{eqn:energy_average} \& \ref{eqn:density_average}.

In conclusion, the convexity condition of the total energy of a finite electronic system with respect to electron count, within density-functional theory, is derived from using the infinite-separation-limit technique based on three minimal,  
conditions, with  sufficient
conditions adapted also
for  approximate DFT. For the reasons discussed, 
this does not imply that the total energy 
is always convex at the ground-state for
all DFT formulations, and counterexamples can
be found (for example) in pure-state DFT. 
Our result nonetheless confirms a stringent constraint on the exact exchange-correlation functional. This proof also
lifts an outstanding assumption in the proof of the piecewise linearity condition with respect to electron count, which
is central to understanding the Kohn-Sham band-gap and the exchange-correlation derivative discontinuity of DFT.

The research conducted in this publication was funded by the Irish Research Council under grant number GOIPG/2020/1454. EL gratefully acknowledges financial support from the Swiss National Science Foundation (SNSF -- project number 213082).

\bibliography{main}

\end{document}